\documentclass[aps,prb,twocolumn,showpacs,amsmath,amssymb,superscriptaddress]{revtex4}
\usepackage{graphicx,dcolumn,bm,epsfig,epstopdf,comment,color}

\begin{document}
%\preprint{APS/123-QED}
\title{The magneto-structural relationship in the tetrahedral spin chain oxide CsCoO$_2$}
\author{N. Z. Ali}
\affiliation{Max-Planck-Institut f\"{u}r Festk\"{o}rperforschung,
  Heisenbergstr.\ 1, D-70569 Stuttgart, Germany}
 \author{R. C. Williams}
 \email{r.c.williams@durham.ac.uk}
  \author{F. Xiao}
  \author{S. J. Clark}
  \author{T. Lancaster}
\affiliation{Durham University, Department of Physics, South Road,
  Durham, DH1 3LE, UK}
\author{S. J. Blundell}
\affiliation{Oxford University Department of Physics, Parks Road, Oxford, OX1 3PU, UK}
\author{D. V. Sheptyakov}
\affiliation{Laboratory for Neutron Scattering and Imaging, Paul Scherrer Institut, CH-5232 Villigen, Switzerland}
\author{M. Jansen}
\affiliation{Max-Planck-Institut f\"{u}r Festk\"{o}rperforschung,
  Heisenbergstr.\ 1, D-70569 Stuttgart, Germany}  

\date{\today}

\begin{abstract}
We have investigated the structural and magnetic transitions in CsCoO$_2$ using calorimetric measurements, neutron powder diffraction (NPD), density functional theory (DFT) calculations and muon-spin relaxation ($\mu$SR) measurements. 
CsCoO$_2$ exhibits three-dimensional long-range antiferromagnetic (AFM) order at 424~K, resulting in antiferromagnetic alignment of chains of ferromagnetically ordered Co-Co spin dimers. Although there is no change in magnetic structure around a structural transition at $T^{*}=100$~K, the resulting bifurcation of corner-shared Co--O--Co bond angles causes a weakening of the AFM interaction for one set of bonds along the chains. Consequently, the system undergoes a complex freezing out of relaxation processes on cooling.

%In the adjacent CoO$_2$ layers, the long range magnetic ordering repeats, being naturally offset by a $(0.5,0.5,0)$ translation. 
%High intensity NPD datasets were collected allowing the refinement of all magnetic and crystal structural parameters of both low-temperature $C2/c$ ($\alpha$) and high-temperature $Cmca$ ($\beta$) phases of CsCoO$_2$. At the displacive structural phase transition ($T^{\star} \approx 100~\mathrm{K}$), the nuclear intensities follow the trend of lowering the symmetry accompanied by monoclinic distortion with subsequent splitting of peaks in the patterns of the $\alpha$ phase. The magnetic structures are essentially the same for both the $\alpha$ and $\beta$ phases. We find that the direction of the Co$^{3+}$ magnetic moments is along the $c$-axis of the crystal lattice, with respective ordered magnetic moments of 3.089 ($2~\mathrm{K}$) and 2.628 ($300~\mathrm{K}$) $\mu_{\mathrm{B}}$/Co$^{3+}$ for $\alpha$- and $\beta$-CsCoO$_2$, respectively. 
%Density functional theory calculations support the proposed magnetic structure comprising strongly ferromagnetic intradimer interactions [$J = 511(1) ~\mathrm{K}$], with weaker antiferromagnetic interactions along the cobalt chains in the crystallographic $a$-axis [$J' = -39(1)~\mathrm{K}$].
%The emergence of single-frequency oscillations in the zero-field $\mu$SR asymmetry spectra is indicative of quasi-static long-range magnetic ordering below the remarkably high N\'{e}el temperature $T_{\mathrm{N}} = 424(2)~\mathrm{K}$. 

\end{abstract}
\maketitle

Cobalt-based multinary oxides continue to attract significant interest due to their complex phase diagrams and the richness of their magnetic and electronic properties, induced by strong correlation between spin, charge, and orbital degrees of freedom. Together with chemical composition and synthesis conditions, the spin states in oxocobaltates are sensitive to external physical parameters such as temperature and pressure. A subtle energy balance between the crystal field splitting and Hund's rule exchange energy in the $3d$ states ultimately regulates the spin state of individual Co ions, offering new potential for tuning materials properties.\cite{maignan,goodenough,pouchard}
The most commonly encountered coordination geometry in multinary oxocobaltates(III) is octahedral, resulting in the low-spin $3d^6$ electron configuration ($S=0$; $t_{2g}^6e_g^0$). %Prominent representatives, where edge-sharing distorted CoO$_6$ octahedra build up sheets, separated by layers of lighter alkali cations, feature diamagnetic ground states (LiCoO$_2$,\cite{bongers,mizushima}) more or less pronouncedly superimposed by temperature independent paramagnetism, as encountered for example in NaCoO$_2$ or AgCoO$_2$.\cite{jansen74,stahlin} 
The less commonly observed tetrahedral coordination geometry, with the weaker crystal field splitting, favours the high-spin state ($S=2$; $e_g^3t_{2g}^3$).\cite{jansen75,delmas,sofin,birx,stusser} 
%Alkalioxocobaltates(III) containing tetrahedrally coordinated Co$^{3+}$ encompass a broad spectrum of interconnected tetrahedral linkages. Isolated CoO$_4$ tetrahedra may serve as basic structural units, as in Na$_5$CoO$_4$\cite{sofin} and Li$_3$Na$_2$CoO$_4$.\cite{birx} Knitting up, we encounter two edge-sharing CoO$_4$ tetrahedra in Na$_6$Co$_2$O$_6$, containing the novel Co$_2$O$_6^{6-}$ anion. Interestingly, attaching to the Co$_2$O$_6$ unit two  CoO$_3$ triangles has resulted in the unique oligomeric Co$_4$O$_{10}^{10-}$ anion in Na$_{10}$Co${_4}$O$_{10}$.\cite{stusser} Finally, 
Within the {\emph A}MO$_2$ family of ternary oxides, the recently discovered CsCoO$_2$ features a singular crystal structure.\cite{ali} Here Co$^{3+}$ is again in a tetrahedral coordination
with edge-linked
CoO$_{4}$ tetrahedra
  forming `bow-tie' shaped Co$_2$O$_6^{6-}$ dimers which are then connected via corner-shared oxygen ions to form spin chains, creating an overall butterfly motif.
% Edge-sharing tetrahedra were previously reported to be unfavourable for such oxides due to repulsion of the tetrahedrally coordinated cations across the shared edge.\cite{gal} The resulting two-dimensionally extended polyoxyanion layers are separated by cesium ions, and adjacent layers are naturally offset by a $(0.5, 0.5, 0)$ translation.
%In addition to the novelty of this structure, it may also be expected to exhibit magnetic behaviour in accordance with the

\begin{figure}[t]
\centering
\includegraphics[width=7cm]{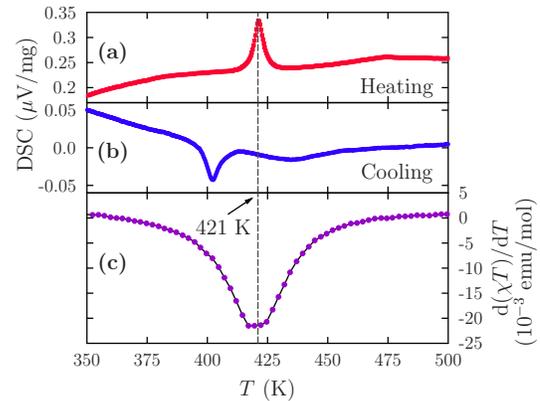}
\caption{\label{fig:thermal}  Peaks within the heating (a) and cooling (b) curves in the DSC measurements indicate an enthalpy of transition $\Delta H$ associated with the magnetic phase transition around $T \approx 421~\mathrm{K}$. (c) The feature in the Fisher heat capacity at $T_{\mathrm{N}} \approx421~\mathrm{K}$ verifies the magnetic origin of the transition.}
\end{figure}

The special connectivity engendered by this architecture
leads to very strongly coupled antiferromagnetism.
According to the  Goodenough-Kanamori-Anderson (GKA) rules %, within each  $_\infty ^{~2}$CoO$_2$  polyoxyanion layer of tetrahedral units. For 
the interdimer superexchange interaction (mediated via corner-shared oxygen ions, denoted O2) the  Co--O2--Co bonding angle between neighboring Co atoms lies close to 180$^{\circ}$; hence the subsequent  interaction between partially filled $d$ orbitals is strongly anitferromagnetic (AFM). Conversely, intradimer Co--Co neighbors are coupled through the intervening edge-shared oxygen ions (denoted O1) with bonding angles close to 90$^{\circ}$, and hence the superexchange interaction is expected to be ferromagnetic (FM).
Measurements of the magnetic
susceptibility \cite{ali}  show that the consequence of the 
connectivity between Co$^{3+}$ ions in such a structure is canted antiferromagnetic
order below the remarkably high temperature of
$T_{\mathrm{N}}=430$~K. 
Above the ordering temperature the
susceptibility becomes temperature independent, suggesting the
persistence of very strong exchange coupling. Despite the progress made in understanding this system, the
magnetic behaviour that accompanies a structural transition that
occurs around 100~K (hereafter denoted $T^{\star}$), involving a monoclinic distortion
resulting from a tilting of the
edge-sharing tetrahedra, remains poorly understood. 
We have therefore carried out a detailed investigation into the magnetism of this system, whose aim is to determine the low energy model of spin interactions and 
to elucidate both the nature of the ordering transition and the magnetic effects that result from the structural transition. 

%In retrospect, the issue of timescale is found to be of crucial importance in many magnetic systems particularly those with high-spin Co$^{3+}$ ions in a tetrahedral coordination. 
%In order to shed light on the complex magnetic exchange patterns present in CsCoO$_2$, we have undertaken the maiden local probe study of this unusual system in order to elucidate the nature of its static and dynamic magnetic properties across the wide temperature range of interest, employing neutron powder diffraction (NPD) and muon spin relaxation ($\mu$SR) experiments on polycrystalline powder samples.

%Despite the progress made in understanding this system, aspects of its magnetism remain perplexing.

Thermodynamic measurements
confirm the presence of the reported magnetic transition around  $T_{\mathrm{N}}=424 ~\mathrm{K}$.\cite{ali} 
The results of differential scanning calorimetry (DSC \cite{SI}) and the ‘Fisher heat capacity method’,\cite{fisher} are shown in Fig.~\ref{fig:thermal}. 
The DSC thermogram measured on heating displays an endothermic peak centered around $421 ~\mathrm{K}$ corresponding to the onset of long range AFM ordering, in good agreement with the N\'{e}el temperature observed by other methods (see below). The rate of temperature change only slightly affects the enthalpy change but does not alter the transition temperature range. There is a substantial change in magnetic entropy and large thermal hysteresis ($\Delta T \approx 15 ~\mathrm{K}$) at $T_{\mathrm{N}}$ between the heating and cooling scans of the DSC plot. In addition, employing the Fisher heat capacity method we have calculated $d(\chi T)/dT$ {\it vs.} $T$ from our magnetic susceptibility data\cite{ali} as shown in Fig. \ref{fig:thermal}(c), displaying a pronounced minimum consistent with the onset of long range AFM ordering. 
%validating the solely magnetic origin of the transition, without any structural variations involved, at  $T_{\mathrm{N}}$.

\begin{figure}[t]
\centering
\includegraphics[width=6cm]{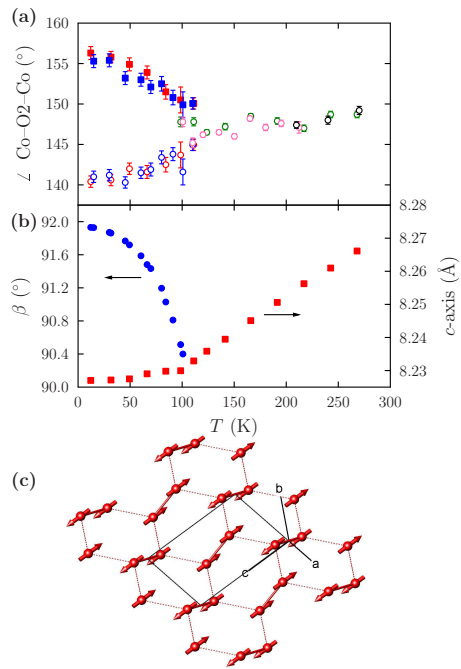}
\caption{\label{fig:neutron} Behavior of the NPD refined nuclear structure for CsCoO$_2$ around the structural phase transition ($T^{\star} \approx 100~\mathrm{K}$): (a) Average Co-O2-Co bond angle involving corner-shared oxygen ions, (b) $c$-axis lattice constant and the monoclinic angle $\beta$ within $\alpha$-CsCoO$_2$. (c) Magnetic structure of a single CoO$_2$ layer within $\alpha$-CsCoO$_2$.}
\end{figure}

In order to elucidate the magnetic structure below $T_{\mathrm{N}}$ and the details of the structural transition around $T^{\star}$, neutron powder diffraction (NPD) measurements were performed. 
The NPD patterns collected between $100 \-- 700 ~\mathrm{K}$ are consistent with orthorhombic $Cmca$ symmetry and can be correctly fitted employing the structural model proposed in our earlier work.\cite{ali} Below $T^{\star} \approx 100 ~\mathrm{K}$ a displacive structural phase transition from the high temperature $Cmca$ space group (hereafter the $\beta$ phase) to monoclinic $C2/c$ space group (hereafter the $\alpha$ phase) is observed as the temperature is reduced, producing a set of new reflections and splitting of certain nuclear intensities accompanying the lowering of symmetry.
The structural phase transition is apparent in the temperature dependence of the monoclinic angle $\beta$, and is accompanied by a change in slope in the $T$-dependence of the $c$-parameter of the unit cell as depicted in Fig. \ref{fig:neutron}(b). While there are no observable features in the temperature dependences of the Co--O average distance, and of the Co--O1--Co bond angles via the edges of the CoO$_4$ tetrahedra, the Co--O2--Co bond angles via the corners of the CoO$_4$ tetrahedra also reveal the transition: from two distinct angles in the monoclinic $\alpha$ phase, a single one emerges upon entering the orthorhombic $\beta$ phase, with a value close to the mean of the two low-$T$ angles [Fig. \ref{fig:neutron}(a)]. This transition is sharp,
but results in no discontinuities in the refined values of the bond angles; they approach each other in a gradual manner with increasing temperature. 
The refinement of the crystal structures above and below $T_{\mathrm{N}} \approx 430 ~\mathrm{K}$ (within the paramagnetic and magnetically ordered phases respectively) reveals no significant modification in the nuclear structure, indicating this transition is of solely magnetic origin. 

At temperatures below $440 ~\mathrm{K}$, in both the $\alpha$ and $\beta$ phases, the NPD patterns of CsCoO$_2$ also contain magnetic diffraction peaks due to the long range order (LRO) of the magnetic Co$^{3+}$ ion spins.
At the structural phase boundary ($T^{\star} \approx 100 ~\mathrm{K}$), the extra (magnetic) intensities follow the trend of lowering the symmetry, for instance the magnetic intensity contained in the position  of the $(1,1,1)$ peak of the crystal structure in the $\beta$ phase is split into the extra intensity in the $(-1,1,1)$ and $(1,1,1)$ peaks in the patterns of the $\alpha$ phase below $T^{\star}$, in a similar fashion to the nuclear intensities. However, no magnetic diffraction pattern changes are observed upon cooling below the structural transition, indicating that the magnetic ordering type and pattern is common to both structural phases. 
(It is possible that this temperature independence of the average Co--O bond distances and of the Co--O--Co bond angles via the edge-shared O1 oxygen atoms which makes the magnetic ordering so robust against the structural phase transition occurring at $T^{\star}$.)
All the magnetic diffraction peaks have been indexed with the propagation vector ${\bm k} =(0,0,0)$, for both the $\alpha$ and $\beta$ phases. The symmetry analysis for this propagation vector and Co ion locations within both the $C2/c$ and $Cmca$ symmetries, has been carried out with the program SARA{\it h}~-2K,\cite{wills} and all symmetry-reasonable magnetic ordering schemes have been verified against the Rietveld refinements. For both the $\beta$ orthorhombic and the $\alpha$ monoclinic phase, just one irreducible representation is compatible with the magnetic ordering model, satisfactorily explaining the observed magnetic intensity patterns. The magnetic ordering is essentially identical for both the $\alpha$ and $\beta$ phases and is illustrated in Fig. \ref{fig:neutron}(c) for one buckled CoO$_2$ layer. 
This spin order consists of  ferromagnetically ordered Co-Co dimers, which are themselves
 antiferromagnetic ordered. 
The intradimer Co-Co interatomic distance is significantly shorter than the corresponding interdimer Co-Co junctions. In the directions of the closest interdimer distances in the $ac$-planes (along the diagonal $\bm{a+c}$ and $\bm{a-c}$ directions), the ordering between the Co moments in the adjacent dimers is antiferromagnetic, thus leading to the overall compensation of the total magnetization. In the adjacent CoO$_2$ layer, the magnetic ordering repeats, being naturally offset by a $(0.5, 0.5, 0)$ translation. 
We find that the direction of the Co$^{3+}$ magnetic moments is along the $c$-axis of the crystal lattice. The irreducible representations leading to this magnetic ordering type are $\Gamma_1$ and $\Gamma_5$ 
(in the notation of SARA{\it h}~-2K) 
for the $\alpha$ monoclinic and $\beta$ orthorhombic phases of CsCoO$_2$ correspondingly, which in principle do not preclude Co magnetic moment components along the other crystal axes. In the $\alpha$ phase, an admixture of the $a$-component with identical ordering type as for the actual $c$-direction could be possible, while for both the $\alpha$ and $\beta$ phases, an admixture of the ferromagnetic ordering with the $b$-component could also be symmetrically reasonable. 
%We have carefully checked our refinements and may conclude that within our data statistics, 
We are unable to assign any significantly meaningful values to the magnitudes of these two ordering types. This means that at all temperatures $T \leq T_{\mathrm{N}}$ where CsCoO$_2$ exhibits magnetic LRO, the only definitely confirmed moment direction is along the $c$-axis of the unit cell. We have determined unusually low magnitudes for the ordered magnetic moments of $3.089 ~(2 ~\mathrm{K})$ and $2.628 ~(300 ~\mathrm{K}) \mu_{\mathrm{B}}$/Co$^{3+}$ for $\alpha$- and $\beta$-CsCoO$_2$, respectively. The suppression of the Co$^{3+}$  magnetic moment, as compared to the predicted spin-only value of $4\mu_{\mathrm{B}}$, can be attributed to strong Co($3d$)—O($2p$) hybridization.

\begin{figure}[t]
\centering
\includegraphics[width=7cm]{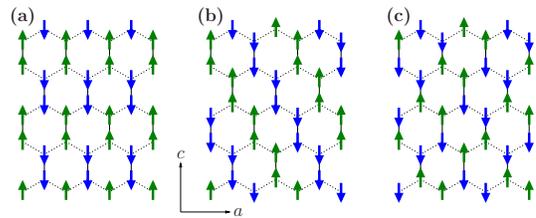}
\caption{\label{fig:struc} (a) Experimentally observed spin structure formed by the Co ions. (b,c) Example higher energy spin structures. The Co ions interact via edge-shared oxygen intradimer (solid line) and corner-shared interdimer (dashed line) superexchange pathways. (See main text.)}
\end{figure}

To further resolve a microscopic picture of the possible static magnetic structures of CsCoO$_2$, we performed a sequence of density functional theory (DFT) total energy calculations using the {\sc castep} code.\cite{clark,hasnip,SI} In agreement with the NPD results, the lowest energy configuration for both the $\alpha$ and $\beta$ phases was found to have strongly ferromagnetically aligned Co-Co dimers, with weaker antiferromagnetic order along the Co chains parallel to the crystallographic $a$-axis [Fig. \ref{fig:struc}(a)]. The weak AFM  interdimer interaction can be easily flipped, as illustrated in Fig.~\ref{fig:struc}(b), at an energy cost of $\Delta E\approx0.034~\mathrm{eV}$/``bond'' (for the low-$T$, $\alpha$ phase). This energy cost may be related to the exchange interaction strength $J$ within the Heisenberg Hamiltonian term for a single exchange bond ${\cal H}=-J \bm{ S_1 \cdot S_2}$ (employing the single-$J$ convention) via $\Delta E = (1/2) J S_{\uparrow \uparrow}(S_{\uparrow \uparrow}+1)$, where $S_{\uparrow \uparrow}=2S$ is the spin quantum number of the dimer triplet state. For Co$^{3+}$ in the high-spin state, $S=2$ and therefore $J=\Delta E /10$. Hence the exchange constant for the weak AFM pathway along the corner-shared O$^{2-}$ ion is predicted to be $J'=-39(1)~\mathrm{K}$. The ferromagnetic intradimer coupling (via the edge-shared oxygen superexchange pathway) is far stronger. The triplet-singlet energy difference between these Co spins was found to be $\Delta E\approx0.44~\mathrm{eV}$, corresponding to an exchange strength of $J=511(1)~\mathrm{K}$ [Fig. \ref{fig:struc}(c)].
There are a number of possible AFM states  analogous to Fig. \ref{fig:struc}(b) (which preserve the FM intradimer spin configuration) that are close in energy and likely to be populated via thermal fluctuations (see below). 
%This suggests that the anomalous behaviour observed in the muon precession frequency $\nu$ might be attributable to the system exploring these energetically similar magnetic states, %comprising flipped dimers, in the $100 \lesssim T \lesssim 200~\mathrm{K}$ region. 
Ordered moment sizes (constrained to be collinear) were found to be $3.25\mu_{\mathrm{B}}$ per Co ion, with some spin density transferred onto the edge-shared oxygen ions ($1.2\mu_{\mathrm{B}}$ per ion), in good agreement with the low-$T$ value obtained through NPD refinements.

In order to gain insight into the unusually large intradimer exchange coupling, we have carried out Mulliken analysis by projecting the valence electron wave functions onto an atomic orbital basis set. Our main finding is that, although the corner-shared oxygen sites carry no net spin,  the edge-shared oxygens (which link Co ions within a dimer) are significantly spin polarised to $\pm 0.28 \hbar$. This is indicative of an enhanced superexchange coupling within the dimers, with one oxygen spin channel having an effective configuration $2p^{3}$ while the opposite spin channel has $2p^{2}$.

\begin{figure*}[ht!]
\centering
\includegraphics[width=14cm]{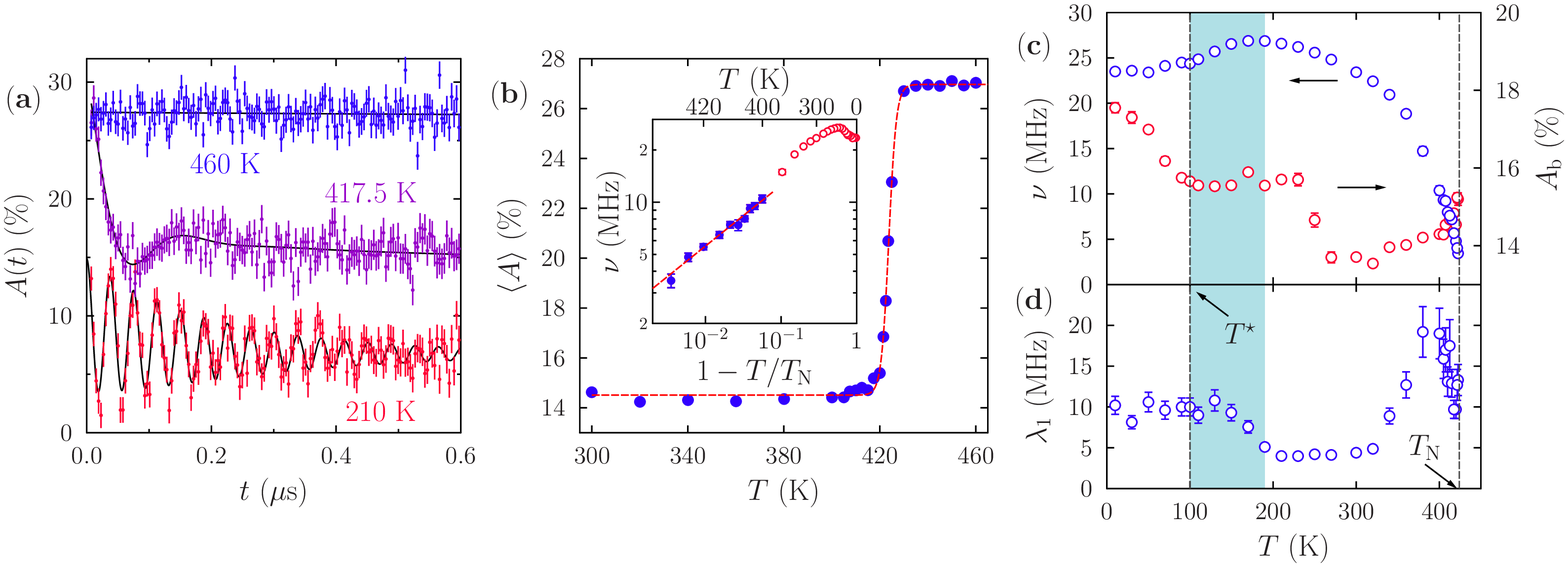}
\caption{\label{fig:muon}(a) $A(t)$ spectra, showing single-frequency oscillations in the ordered regime with a fit described in the text (the $210 ~\mathrm{K}$ data are offset by $10\%$ for clarity). (b) Time-averaged asymmetry near the critical region, displaying an abrupt drop in asymmetry $\Delta A = A_2-A_1$. The dashed line is the fit described in the text. Inset: scaling plot for the order parameter $\nu$. The linear fit indicates the region where the behaviour is critical. (c) Fitted parameters to the oscillatory relaxation function for all asymmetry data ($t \leq 9.5~\mathrm{\mu s}$) for temperatures $T< T_{\mathrm{N}}$; frequency $\nu$ and  baseline asymmetry $A_{\mathrm{b}}$ and (d) transverse relaxation rate $\lambda_1$. The N\'{e}el temperature $T_{\mathrm{N}}=424~\mathrm{K}$ and the anomalous temperature region $100\lesssim T \lesssim 190~\mathrm{K}$ are indicated. }
\end{figure*}

%Muons are a sensitive local probe of both static and dynamic magnetism,\cite{blundell99} and therefore $\mu$SR measurements were made on the GPS instrument at S$\mu$S. Observing the time-dependence of the muon polarization (via the positron asymmetry\cite{SI}) allows us to study the critical behaviour near $T_{\mathrm{N}}$, dynamics of magnetic fluctuations and interplay between the magnetic and crystallographic structures around $T^{\star}$.

To probe the local magnetism on a timescale distinct from that measured by NPD, muon-spin relaxation ($\mu$SR) measurements \cite{blundell99} were made on CsCoO$_{2}$.
Spontaneous oscillations in the asymmetry spectra measured in zero-field are clearly visible for temperatures below around 420 K [Fig. \ref{fig:muon}(a)]. This constitutes unambiguous evidence of quasi-static magnetic LRO throughout the bulk of the material. A clear indication of the transition temperature is given by considering the time-averaged ($t \leq 9.5~\mu \mathrm{s}$) asymmetry data, which drops abruptly upon cooling, where asymmetry is lost as the material enters the magnetically ordered state. Fig. \ref{fig:muon}(b) shows this drop in average asymmetry, which may be fitted with a Fermi-like step function
%\begin{equation} \label{FD}
$\langle A  \rangle (T) = A_2 + \frac{A_1-A_2}{e^{(T-T_{\mathrm{c}})/w} + 1},$
%\end{equation}
providing a method for extracting the transition temperature\cite{steele} by parametrising the continuous step from high- (low-) $T$ asymmetry $A_2$ ($A_1$) with mid-point $T_{\mathrm{c}}$ and width $w$. The fit yielded values of $T_{\mathrm{c}}=423.7(2)~\mathrm{K}$ and $w=1.57(2)~\mathrm{K}$, in agreement with the value previously obtained.\cite{ali}

Asymmetry specta below $T_{\mathrm{N}}$ were best fitted with the single-frequency oscillatory relaxation function $A(t)= A_1\cos(2\pi \nu t)e^{-\lambda_1t}+A_2e^{-\lambda_2t}+A_{\mathrm{b}}$, where the oscillating amplitude $A_1$ was fixed to its average value of $5.7\%$. The non-relaxing baseline contribution $A_{\mathrm{b}}$ is attributable to muons which stop in the Ti sample holder, and, more importantly, to the non-precessing component of muon spins which lie parallel to the local magnetic field. Parameters resulting from this fit are displayed in Fig. \ref{fig:muon}(c,d).

 Upon cooling below $T_{\mathrm{N}}$, the precession frequency $\nu$ increases in the expected manner, and the relaxation rates $\lambda_i$ peak, as is typical for an AFM phase transition (Fig.~\ref{fig:muon}). 
Frequencies in the critical region $400~\mathrm{K}\leq T < T_{\mathrm{N}}$ were fitted to 
$\nu (T) = \nu (0) \left( 1 - T/T_{\mathrm{N}} \right) ^{\beta}$, 
where $T_{\mathrm{N}}=424~\mathrm{K}$ was fixed [see Fig. \ref{fig:muon}(b) inset]. The fit yielded a value for the critical parameter $\beta = 0.35(2)$,\cite{note} which is consistent with that expected for a 3D Heisenberg antiferromagnet. %Values of $\beta$ and $T_{\mathrm{N}}$ remained unchanged (to within uncertainties) when the critical temperature was also allowed to vary within the fit. 
Moreover, dipole field simulations\cite{SI} were performed on CsCoO$_2$ using the proposed magnetic spin structure and crystal parameters obtained from x-ray diffraction at $50 ~\mathrm{K}$ and $296 ~\mathrm{K}$.\cite{ali} A Bayesian analysis of this calculation \cite{blundell12,SI} shows that the observed muon precession frequencies are fully consistent with a 
moment size of $2.63\mu_{\mathrm{B}}$ from the neutron measurement and further suggests  a muon stopping site approximately $1$~\AA{} from the corner-shared oxygen ions. 

In addition to the behavior observed around $T_{\mathrm{N}}$ described above, which is quite typical for an AFM transition, we find that on cooling though the region $T < T_{\mathrm{N}} $, there are a number of additional, notable features in the $\mu$SR data.
We find that on cooling below $\approx190~\mathrm{K}$ the muon precession frequency $\nu$ is smoothly suppressed [Fig. \ref{fig:muon}(c)]. 
%No discontinuities are observed in the fitted parameters around this temperature, consistent with the picture of a broad, virtually second order structural transition. 
This is accompanied by a steady increase in $\lambda_1$, such that $\lambda_1$ is larger by a factor of about 2 below $T ^{\star}$ than in the region $200~\mathrm{K} \lesssim T \lesssim 300~\mathrm{K}$ [Fig. \ref{fig:muon}(d)]. Below $T^{\star} \approx 100~\mathrm{K}$, $\lambda_1$ and $\nu$ both level off at constant values. 
The decrease of $\nu$ below 200~K indicates a reduced value of average magnetic field strength experienced at the muon stopping sites. 
In the fast fluctuation limit we expect that the relaxation rate $\lambda \propto \langle (B-\langle B \rangle)^2 \rangle \tau $ (i.e.\ the second moment of the magnetic field distribution multiplied by the correlation time $\tau$ \cite{hayano})  and so the additional dephasing indicates a broadening of the distribution of static magnetic field strengths experienced by the muon ensemble, or an increased correlation time as relaxation channels freeze out on the muon time scale. 
It is possible that this behaviour reflects the 
the system exploring some of the energetically similar magnetic states predicted by the DFT calculations described above. 
These states are realized by  flipping the overall spin of a dimer, while preserving their FM intradimer spin configuration [Fig. \ref{fig:struc}(b)]. Although this could conceivably lead the muon ensemble to experience a broader static magnetic field distribution, with a lower mean field strength, one would expect this to occur below $T^{\star}$ rather than above it. Moreover, the  refined cobalt moment sizes from NPD data (shown in the SI\cite{SI}) do not show this suppression below $200~\mathrm{K}$, suggesting that the effect could be dynamic and related to the issue of timescale. Such effects are therefore not seen in the neutron measurements as these effectively take a `snapshot' of the spin distribution, when compared to the muon GHz timescale.

Additional evidence for the influence of dynamics comes from the baseline asymmetry $A_{\mathrm{b}}$, which increases abruptly at around $250~\mathrm{K}$ and upon further cooling below $T^{\star}$ gradually increases further [Fig. \ref{fig:muon}(c)]. Since relaxation of those muon spins that initially point along the direction of the local magnetic field can only be achieved by dynamic relaxation precesses, the increase of such a non-relaxing signal is usually indicative of a freezing of relaxation processes. We therefore have evidence for an initial, abrupt freezing of some relaxation channels around 250~K, prefiguring the suppression of the precession frequency $\nu$, with a further, smooth increase below 100~K that seems to track 
%followed by a further, smooth increase that seems to be reminiscent of 
the size of the monoclinic distortion as indicated by both the change in unit cell angle $\beta$ and the splitting of the corner-shared bond angle Co--O2--Co (Fig.~\ref{fig:neutron}). 
This monoclinic distortion leads to a disproportionation of the corner-shared oxygen bond angles; above $T^{\star}$ there is one unique value of about $150^{\circ}$, whereas below there appear two classes of bond with angles of around $140^{\circ}$ and $155^{\circ}$.\cite{SI} These values lie within the linear GKA rule regime, but not particularly close to the fully linear angle of $180^{\circ}$ (which leads to strong AFM coupling), hence the weaker AFM exchange interaction strength $J'$. Meanwhile the edge-shared intradimer bond angle only slightly deviates from $90^{\circ}$ both above and below $T^{\star}$ ($88^{\circ}$ and $86^{\circ}$ respectively\cite{SI}) and therefore displays a strong FM superexchange interaction as expected from GKA. The weaker AFM interdimer coupling drives the transition to LRO, and so the system would be sensitive to the bifurcation of the corner-shared oxygen bond angle below $T^{\star}$, where the AFM exchange interaction strength $J'$ would also split into two unequal values. When these superexchange pathways become inequivalent in the monoclinic phase it is possible that this allows some relaxation channels to freeze out, leading to a greater static component of the magnetic field, and hence the increase in non-relaxing asymmetry $A_{\mathrm{b}}$ observed.

In conclusion, neutron powder diffraction has enabled the determination of the microscopic magnetic structure of CsCoO$_2$ below the N\'{e}el temperature. %, determined to be $T_{\mathrm{N}}=424(2)~\mathrm{K}$. 
The tetrahedrally coordinated high-spin ($S=2$) Co$^{3+}$ ions form a spin configuration comprising strongly FM linked dimers, with weaker AFM interdimer superexchange interactions creating the extended LRO within the extended CoO$_2$ planes. This spin configuration is observed both above and below a structural phase transition around $T^{\star} \approx 100~\mathrm{K}$, and is supported by DFT calculations. %$\mu$SR experiments allowed the critical behaviour near $T_{\mathrm{N}}$ to be examined, and have shed light upon the system's magnetic behaviour around the structural transition. 
The bifurcation of bond angles, and hence superexchange pathway interaction strengths pertaining to the weaker AFM interdimer interactions, result in a complicated freezing of relaxation processes on the muon timescale. 

Part of this work was carried out at the Swiss spallation neutron source SINQ and Swiss muon source S$\mu$S, Paul Scherrer Institut, Villigen, Switzerland. We are grateful for the provision of beamtime, and to Alex Amato for muon experimental assistance. We also wish to thank UKCP and the Archer HPC facility for computer time. This work is supported by the EPSRC (UK).


\begin{thebibliography}{xx}
\bibitem{note}
The value of the critical parameter $\beta$ for a given fitted frequency $\nu$ was found to be sensitive to the time range over which the asymmetry spectra were fitted. The quoted value of 0.35 was obtained for fits over early times only ($t \leq 1~\mu \mathrm{s}$), which are most sensitive to the behavior of the oscillations.
\end{thebibliography}
\end{document}